\documentstyle[preprint,epsfig,aps]{revtex} 

\def\beq{\begin{equation}}
\def\eeq{\end{equation}}

\def\mygam{{\alpha}}

\draft
\tighten
\input epsf

\begin{document}

\title{Shortcuts in the fifth dimension}

\author{Robert Caldwell}
\address{ Department of Physics \& Astronomy \\
Dartmouth College \\
6127 Wilder Laboratory\\
Hanover, New Hampshire 03755 USA}

\author{David Langlois}
\address{Institut d'Astrophysique de Paris, \\
Centre National de la Recherche Scientifique,\\
98bis Boulevard Arago, 75014 Paris, France}

\date{\today}

\maketitle


\begin{abstract}

If our Universe is a three-brane embedded in a five-dimensional anti-deSitter
spacetime, in which matter is confined to the brane and gravity inhabits  an
infinite bulk space, then the causal propagation of luminous and gravitational
signals is in general  different.  A gravitational signal traveling between two
points on the brane can take a ``shortcut'' through the bulk, and appear
quicker than a photon traveling between the same two points along a geodesic on
the brane. Similarly, in a given time interval, a gravitational signal can
propagate farther than a luminous signal.  We quantify this effect, and analyze
the impact of these shortcuts through the fifth dimension on cosmology. 

\end{abstract}

\section{Introduction}

The idea that our Universe may be a boundary of a larger spacetime manifold has
triggered an outburst of creative and profound research in particle physics and
cosmology.   The notion of a boundary or brane-world was first made concrete
in  Horava-Witten theory \cite{HW}, an M-theory in which the gauge fields are
confined to a series of fundamental domain walls and gravity inhabits the bulk
space between the walls. Inspired by such M-theory developments, the extra
dimensions have been exploited in a variety of situations, notably in an
explanation of the mass scale hierarchy problem \cite{ADD98}. The
Randall-Sundrum model \cite{RS99B} has demonstrated that extra dimensions need
not be compact or even small, leading to fascinating speculation for cosmology
and experiment.  That is, these extra dimensions are not just the realm of
abstract theory, but may have observable consequences ranging from astrophysics
\cite{ADD99,BHKZ99,HS99} and accelerators \cite{GRW98,MPP98,H98,CHE00} to the
laboratory \cite{gravtesta,gravtestb,gravtestc}.

The brane-world has become a new forum for the investigation of cosmology. 
Numerous studies have explored the dynamics of inflation, or the generation and
evolution of fluctuation spectra in the early Universe brane-world.  A
significant result which has spurred on much work is the analog of the FRW
equation for the cosmic evolution on a three-brane embedded in five-dimensional
anti-deSitter \cite{BDL00,BDEL00}. Although the gravity is Einsteinian, the
backreaction of the curvature at the brane/bulk interface onto the brane causes
the cosmological expansion law to become
\beq
\left({\dot a \over a}\right)^2 = {{\kappa_{(5)}}^4 \over 36} \rho_{brane}^2
+ {\Lambda \over 6} 
\label{nseq}
\eeq
where the five-dimensional Newton's constant and mass scale are related by
$\kappa_{(5)}^2= M_{(5)}^{-3}$ and where the four-dimensional  Planck mass is
given by $M_{pl}^2=M_{(5)}^3\ell$,  $\ell\equiv \sqrt{-6/\Lambda}$ being the 
anti-deSitter radius of curvature.  It has been recognized  \cite{cosmors} that
if the brane  carries, in addition to ordinary matter, a tension $\sigma$ (such
that $\rho_{brane} = \rho + \sigma$) which compensates the  effect of the bulk
cosmological constant $\Lambda$, or more precisely such that $\kappa_{(5)}^4
\sigma^2=-6\Lambda$, the standard expansion law can be recovered in the
late-time limit when $\rho\ll \sigma$. 

Few brane-world studies have considered the simpler, yet deeper issue of
causality. In previous work, Chung and Freese \cite{CF99B} have demonstrated
that a null geodesic passing through an extra dimension can connect points in
the lower-dimension which are causally disconnected with regard to null
geodesics confined to the brane. They further speculated that such null
geodesics could be used to solve the cosmological horizon problem, in place of
inflation. However, they did not consider a realistic spacetime in their
analysis. Next, Ishihara \cite{I00} has shown quite generally that the
condition for the existence of ``causality violating'' null geodesics which
pass through the anti-deSitter bulk is merely the deviation from a pure
tension-like stress-energy tensor.  That is, provided $\rho_{brane} + p_{brane}
> 0$, then the extrinsic curvature bends the brane concave towards the bulk,
allowing for the existence of such null geodesics. These two results serve as
the starting point for our investigation.

In this paper, we reduce the analysis of graviton propagation in  an infinite,
warped  bulk into a practical form. Our principle result is a useful expression
for what we will call the ``gravitational horizon radius'' in contrast to the
standard, photon horizon radius in an FRW spacetime. This result will allow us
to demonstrate that the horizon problem is not so easily solved: although light
is supplanted by the graviton in determining the causal structure of the
brane-world, the effect in a realistic scenario is small. 

\section{The spacetime}

The starting point for our investigation is a five-dimensional spacetime,
analogous to the Randall-Sundrum model, where we take the extra dimension to be
infinite in extent. 

\subsection{The bulk}

In fact, a generalization of the bulk spacetime is Schwarzschild -
anti-deSitter, which we find convenient for this analysis. The metric can be
written as
\beq 
ds^2=- f(R) dT^2 +f(R)^{-1}dR^2+R^2d\Sigma_k^2,
\eeq
where $d\Sigma_k^2$ stands for the metric of maximally symmetric 
three-dimensional spaces ($k=0$ for a flat three-space, $k=1$ for a 
three-sphere, $k=-1$ for a hyperbolic three-space), and with 
\beq
f(R)=k+{R^2\over {\ell}^2}-{\mu\over R^2}.
\eeq
Here, $\ell$ is the constant curvature radius of anti-deSitter and $\mu$ is the
five-dimensional Schwarzschild-like  mass.  We will be interested primarily in
the simplest case $k=\mu=0$. In addition, we take the bulk to be empty. This is
not generally true, as the bulk is typically filled with other fields such as a
supergravity multiplet, as well as other branes, in more realistic models.
Nevertheless, provided these additional elements are minor, {\it e.g.} the
energy due to the additional fields is negligible compared to the negative
cosmological constant, and the additional branes are distant, then our
assumption of an empty bulk should be reasonable.

\subsection{The brane}

We assume that the spacetime of the three-brane is homogeneous and isotropic. 
Therefore, the trajectory of the brane is simply determined by its position in
the fifth dimension, {\it i.e.} by a function $R_b(T)$. In other terms, the
problem of the motion of the brane in the bulk is analogous to the motion of a
particle in a two-dimensional spacetime with coordinates $R$ and $T$.

It is useful to introduce the proper time $t$ for the brane, defined by
\beq
dt^2=f(R_b)dT^2 -{d R_b^2\over f(R_b)}, 
\eeq
so that 
\beq 
dT={\sqrt{f(R_b)+\dot R_b^2}\over f(R_b)} dt\label{cosmictime}
\eeq
where the dot indicates the derivative with respect to $t$. Then, the induced
metric on the brane is simply
\beq
ds_{brane}^2= -dt^2+R_b(t)^2 d\Sigma_k^2,
\eeq
and $R_b(t)\equiv a(t)$ can be identified with the usual cosmological scale
factor in the brane-world.

\subsection{Null geodesics}

\begin{figure}
\begin{center}
\epsfxsize=4.8 in \epsfbox{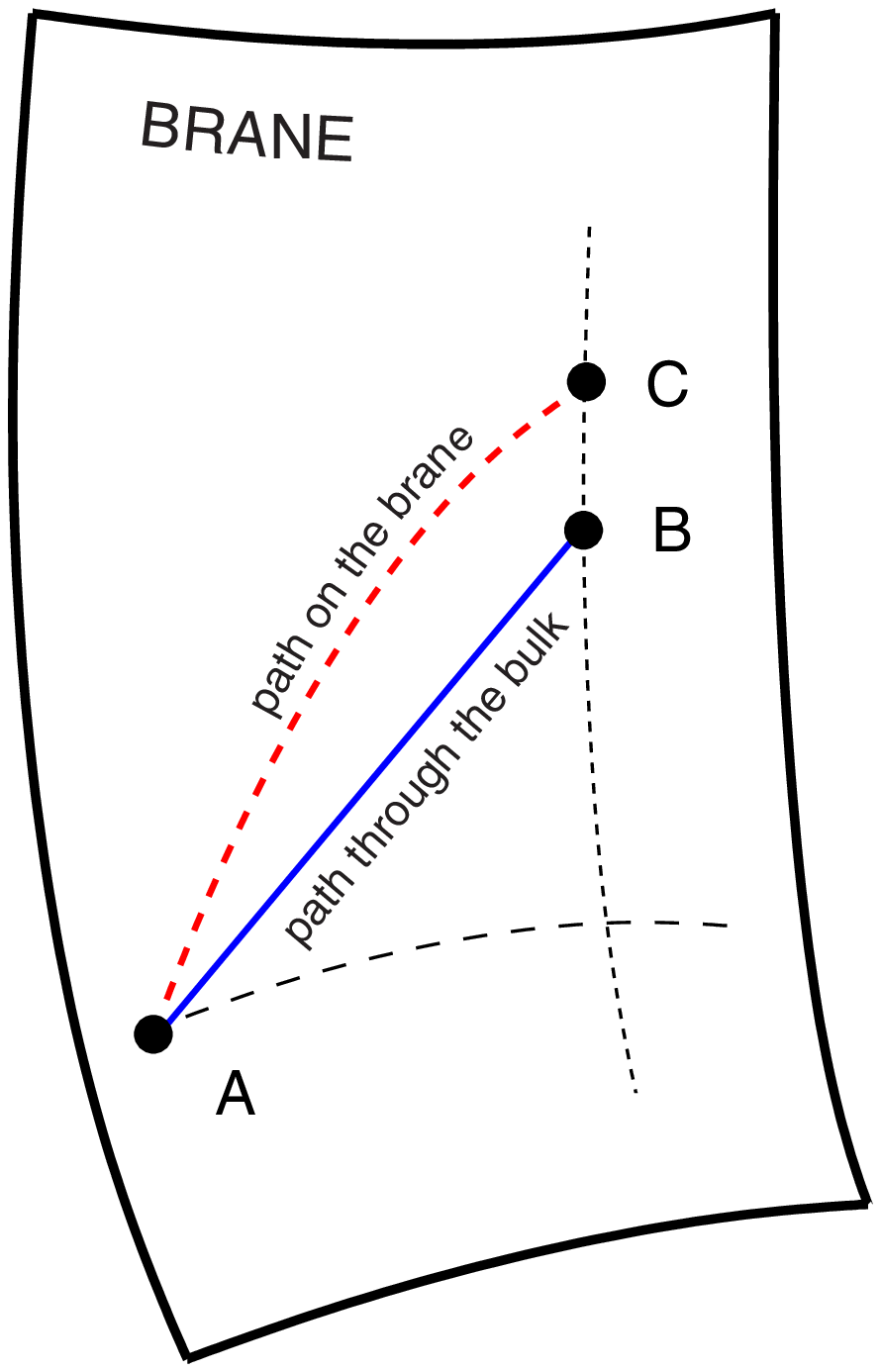}
\end{center}
\caption{The null geodesics on the brane and  through the bulk are represented
schematically.  The line passing from points A to C represents a null geodesic
on the brane, whereas the points A and B are joined by a null geodesic which 
takes a shortcut through the bulk. The light, long dashed line passing through
A represents a hypersurface of fixed cosmological time on the brane; the light,
short dashed line passing through B and C represents the trajectory of points
at a fixed comoving position.  \label{figure1}}
\end{figure}

Our purpose now is to compute the trajectories of null geodesics in the bulk
spacetime, which start from some point within the brane.  Let us consider such
an initial point, A, as illustrated in Figure~\ref{figure1}. It is convenient
to introduce a spherical coordinate system $(r,\theta, \phi)$ in the brane,
which is  centered on A, so that any signal can be described by a radial
geodesic. Then, free to ignore the angular variables $\theta$, $\phi$, we are
left with a three-dimensional problem with a metric
\beq
ds^2=- f(R) dT^2 +f(R)^{-1}dR^2+R^2dr^2.
\eeq
To compute the geodesic trajectories, it is convenient to resort to the 
Killing vectors of the metric, which are here $\left({\partial \over\partial
T}\right)^a$ and $\left({\partial \over\partial r}\right)^a$. If one denotes
$k^a=(dx^a/d\lambda)$ as the vector tangent to the geodesic, then the existence
of these two Killing vectors implies that 
\beq
k_T= -f(R){dT\over d\lambda}=-E \label{kT}
\eeq
and 
\beq
k_r=R^2 {dr\over d\lambda}=P \label{kr}
\eeq
are constants of motion along the geodesics. Imposing moreover that $k^a$ is a
null vector, one finds
\beq
\left({dR\over d\lambda}\right)^2=E^2-P^2{f(R)\over R^2}. \label{Rdot}
\eeq
Combining (\ref{kr}) with (\ref{Rdot}), one easily gets
\beq
\left({E^2\over P^2}-{f\over R^2}\right)^{-1/2}{dR\over R^2}=dr.
\eeq
This is the seed of our result, as it relates distances on the three-brane to
the radial coordinate in the five-dimensional space or equivalently the
expansion scale factor on the brane. In the particular case $k=\mu=0$ it is
straightforward to integrate to get 
\beq
{1\over R_A}-{1\over R}={E\over P}\mygam r,
\eeq
with 
\beq
\mygam \equiv \sqrt{1-{P^2\over E^2l^2}}.
\eeq
Similarly, combining (\ref{kT}) with (\ref{Rdot}), one gets the trajectory of
the geodesic along the time coordinate. The infinitesimal version is 
\beq
{dR\over f\sqrt{1-{P^2 f\over E^2R^2}}}=dT.
\eeq
Once more, in the case $k=\mu=0$ it can be integrated to yield the 
very simple relation
\beq
{1\over R_A}-{1\over R}={\mygam\over {\ell}^2}\left(T-T_A\right).
\eeq
One can also relate directly $T$ to the radius $r$, according to 
\beq
r={P\over E {\ell}^2}\left(T-T_A\right).
\eeq
Finally, it is possible to get rid of the parameters $E$ and $P$ to get the 
following equation for the geodesic
\beq
\left({1\over R_A}-{1\over R}\right)^2+{r^2 \over {\ell}^2}= 
{1 \over {\ell}^4}(T-T_A)^2.
\eeq
Let us now denote B as the point where the null geodesic starting from A again 
crosses the brane. The time difference $T_B-T_A$ can be expressed, using 
(\ref{cosmictime}), in terms of the brane proper time, {\it i.e.} the
brane-world cosmic time:
\beq
T_B-T_A= \ell \int_{t_A}^{t_B} {dt\over a}\sqrt{1+\ell^2H^2}.
\eeq
Then we see that between times $t_A$ and $t_B$, the null geodesic has traversed
a comoving distance $r_g$:
\beq
r_g =   \left( \left[\int_{t_A}^{t_B} {dt\over a}\sqrt{1+\ell^2H^2}\right]^2
- \left[\int_{t_A}^{t_B} {dt\over a} \ell H\right]^2 \right)^{1/2}.
\label{rgeqn}
\eeq
This equation represents the main result of this paper, a simple expression
which gives the horizon radius for the causal propagation of gravitational
signals between two points on the brane through the bulk. Hence, we call this
the {\it gravitational horizon radius}.

The horizon radius for the causal propagation of luminous signals on the brane,
as in the standard FRW cosmology, is given by
\beq
r_\gamma = \int_{t_A}^{t_B} {dt\over a}  
\label{rpeqn}
\eeq
where the subscript indicates that this is the path traveled by photons and
other fields confined to the brane manifold.  We will be interested in cases in
which $r_g$ and $r_\gamma$ are different. Note that, if our universe was
static, i.e. $H=0$,  which in the  present model would correspond to the strict
Randall-Sundrum configuration  \cite{RS99B}, or de Sitter, i.e. $H>0$ and 
constant,   then the photon horizon and the bulk gravitational horizon would be
exactly identical. (This agrees with the results of Ishihara \cite{I00}, since
$\rho_{brane} + p_{brane}=0$.) 

\section{Causal Distances}

There are two interesting regimes for the evaluation of $r_g$, depending on 
the ratio between the Hubble radius and the five-dimensional length scale. 
We examine in turn the two regimes.

\subsection{The low energy regime: $\ell H \ll 1$}

This regime corresponds to a universe governed by the standard FRW equation. 
In this case it is simple to manipulate the integrals in
(\ref{rgeqn},\ref{rpeqn}) using $dt/a = da/(a^2 H)$ to obtain the ratio of the
gravitational to photon distances traveled by a signal propagating between 
times $t_A$ and $t_B$.  Expanding in terms of the small parameter $\ell H$, we
obtain 
\begin{eqnarray}
r_g/r_\gamma &\approx & 1 + \frac{1}{2} \, 
 (\ell H_B)^2 \;{1 + 3 w \over 5 + 3 w} \; 
 \left({a_B \over a_A}\right)^{(5+3w)/2} \cr
&\times &
 \left[ {1 - (a_A/a_B)^{(5+3w)/2} \over 1 - (a_A/a_B)^{(1+3 w)/2}}
 -  {(1 + 3 w)(5+3w) [1 - (a_A/a_B)]^2 \over 
 4 [1 - (a_A/a_B)^{(1+3w)/2}]^2}\,
 \left({a_A \over a_B}\right)^{(1+3w)/2} \right] \cr
&\sim & 
 1 + \frac{1}{2} \, 
 (\ell H_B)^2 \;{1 + 3 w \over 5 + 3 w} \;
 \left({a_B \over a_A}\right)^{(5+3w)/2} 
\end{eqnarray} 
where $w=P/\rho$ is the equation of state of the background  matter on the
brane ({\it e.g.} $w=1/3,\,0$ in the radiation, matter eras),
and the last approximation is valid for $w > -1/3$ and $a_B \gg a_A$. 

Let us consider a signal which would reach us now, at $t_B=t_0$. Then the above
ratio reduces to 
\beq
r_g / r_\gamma \approx 1 + \frac{1}{10} \; (\ell  H_0)^2 \;
\left(1+z\right)^{5/2},
\eeq
where $H_0$ is the present Hubble parameter, and $z$ the redshift of the 
source emitting the signal, which we have assumed to be in the matter-dominated
era. We see that the magnitude of the time delay depends on the curvature
radius, $\ell$, of the 5-dimensional anti-deSitter spacetime. However, based on
precision tests of the gravitational force law, the size of the extra dimension
must be less than $\sim 1$~mm \cite{gravtestc}, so that $\ell H_0\lesssim
10^{-29}$. We conclude that,  although the time delay increases with the
redshift of the source,  it is not enough to compensate for the extremely small
factor  $(\ell  H_0)^2$ in order to obtain a significant cosmological time
delay at present.

\subsection{The high energy regime: $\ell H \gg 1$}

This regime corresponds to the early Universe, for energy densities $\rho
\gtrsim \sigma \approx  M_{Pl}^2/\ell^2 \approx M_{(5)}^6/M_{Pl}^2$.
Interestingly, the leading contribution to the gravitational distance is
independent of $\ell$, so that the ratio becomes
\beq
{r_g \over r_\gamma} \approx \left[ \int_A^B {da \over a^2 H^2} 
\int_A^B {da \over a^2} \right]^{1/2} / \int_A^B {da \over a^2 H}.
\eeq
At these energy scales  the non-standard cosmic evolution of equation
(\ref{nseq}) applies, with $H^2 \propto \rho^2$ ({\it e.g.} radiation, with an
equation of state $w=1/3$ drives $H \propto a^{-4}$). Therefore we find
\beq
{r_g \over r_\gamma} \approx \left[ {(2 + 3 w)^2 \over (5 + 6 w)}
{\left(a_B/a_A - 1\right) \left(1 - (a_A/a_B)^{5+6w}\right)
\over \left(1 - (a_A/a_B)^{2+3w}\right)^2 } \right]^{1/2}
\sim   {2 + 3 w \over \sqrt{5 + 6 w}} \; \sqrt{a_B  \over a_A},
\label{earlyeq}
\eeq
where the last approximation is valid for $w > -2/3$ and $a_B \gg a_A$.
The ratio $r_g /r_\gamma$ thus goes to infinity when $a_A$ goes  to zero.
However, there is a limit to the applicability of this result, since there is a
lower bound on the time for which the physics of this scenario is valid. In
standard cosmology, this limiting time is the Planck time. In a model with
extra-dimensions, the  limiting time is related to the fundamental mass scale
of the theory,  which is here $M_{(5)} = (M_{pl}^2 / \ell)^{1/3}$  (with
$M_{(5)} \sim 10^8$~GeV for $\ell\sim1$~mm).  Indeed, the theory will be
invalid for energy densities  in the brane higher than $M_{(5)}^4$,
which corresponds  to a cut-off Hubble parameter $H\sim M_{(5)}$.  As a result
of this constraint, the largest ratio for the gravitational to luminous horizon
radii is obtained with $t_B\sim \ell$ and $t_A\sim M_{(5)}$, which yields
\beq
{r_g \over r_\gamma} \sim 
\sqrt{a_B \over a_A}\sim \left(H_A\over H_B\right)^{1/8}
\sim \left(M_{(5)}\ell\right)^{1/8}
\sim \left({M_{Pl}\over M_{(5)}}\right)^{1/4}. 
\label{ratio}
\eeq
where we have assumed a non-standard, radiation-dominated era. With the lowest
possible value $M_{(5)} \sim 10^8$~GeV, this gives a maximum ratio
$r_g/r_\gamma\sim 10^3$.

Now we turn to the classic horizon problem.  The ratio of the  horizon radius 
at the present time $t_0$ to the horizon  radius at some early time $t_B$ is
given by
\beq
r_{\gamma 0}/r_{\gamma B} = {\int^{t_0} dt/a \over \int^{t_B} dt/a}
\approx {a_B \over a_0} {H_B \over H_0}.
\eeq
Since this is a standard textbook problem, it is sufficient to observe that
$r_{\gamma 0}/r_{\gamma B} > 1$ is the essence of the horizon problem. For
$t_B\sim \ell$, one finds that
\beq
r_{\gamma 0}/r_{\gamma B} 
\approx {a_B \over a_0} {H_B \over H_0}
\sim \left[ (\ell H_0)^2  (1+z_{eq})\right]^{-1/4}.
\eeq
For $\ell \sim 1$~mm, this gives $r_{\gamma 0}/r_{\gamma B} \sim 10^{14}$. It
is thus clear that the $10^3$ ratio between the bulk gravitational  horizon and
the usual horizon is quite insufficient to account for the  horizon problem.
Even relaxing the bound on $\ell$ due to gravitational  experiments today (by
considering an effective  bulk cosmological constant that varies with time so
that  $\ell$ contracts on millimeter scales after nucleosynthesis) would not
be  sufficient. Indeed, the constraint on $\ell$, or $M_{(5)}$, would then  be
the nucleosynthesis constraint, which can be expressed by the  condition
\beq
\sigma^{1/4} < 1 \;  {\rm MeV},
\eeq
implying a minimum mass $M_{(5)}\sim 10^4$~GeV (since $\sigma\sim
M_{(5)}^6/M_{Pl}^2$). The ratio $r_g/r_\gamma$ given by  (\ref{ratio}) can then
be increased by one order of magnitude up to $10^4$, while the ratio $r_{\gamma
0}/r_{\gamma B}$ can be decreased to the value $10^8$. Still, this is not
enough to solve the horizon problem.

\section{Analysis}

We have shown that shortcuts through the fifth dimension, with the
gravitational horizon radius given by equation (\ref{rgeqn}), are not short
enough to solve the classical horizon problem. We have furthermore argued that
a time evolving bulk energy density, which would permit $\ell$ to start out
large and  decrease with time, cannot fully solve the problem due to other
constraints.

We caution that our results are valid strictly for the case of an empty bulk
spacetime with a single, infinite, extra dimension as described in this paper.
Motivated by the Randall-Sundrum scenario and other work on brane-world
cosmology, this case has added appeal due to the simplicity of the geodesic
paths. Naturally, one may ask how these results apply to more general cases.
Since the shortcut is a consequence of the warping of space in the extra
dimension, there is no shortcut for compact, flat extra dimensions. We have not
explored the case of more than one extra warped dimension (we are unaware of
any such models in cosmology),  though such scenarios, with additional bulk
fields,  might be more realistic in the context of certain particle physics
models. Once the spacetime metric is known, one should repeat the procedure
described in this paper.

\begin{figure}
\begin{center}
\epsfxsize=6.5 in \epsfbox{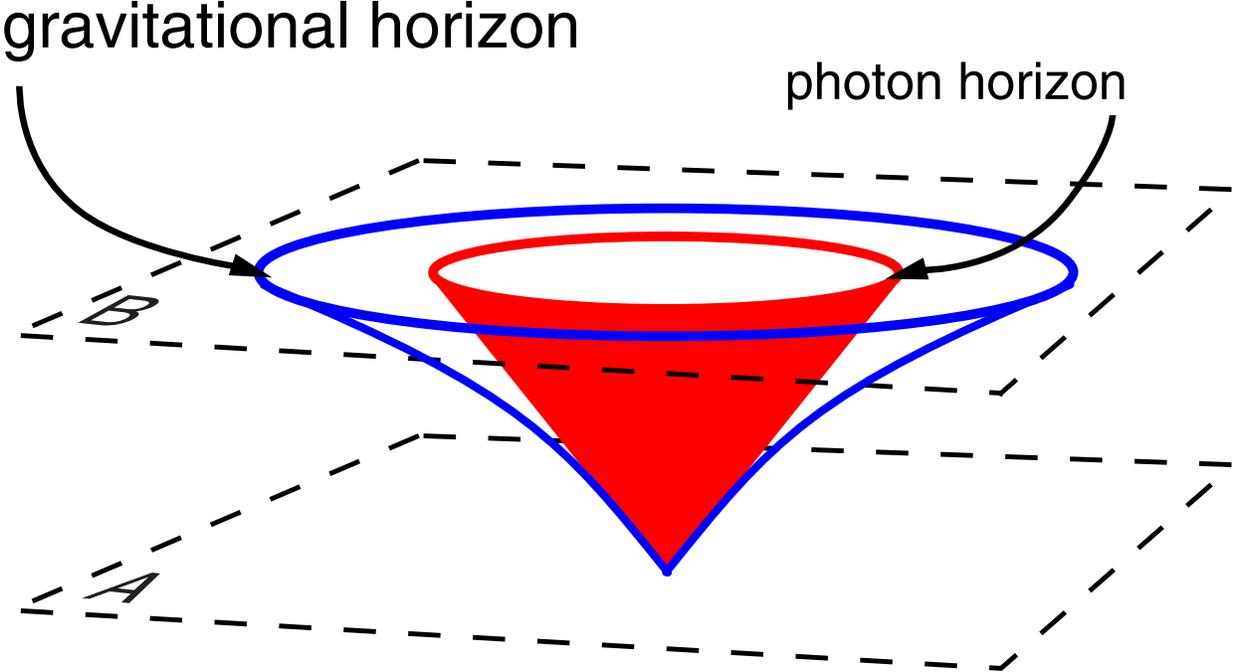}
\end{center}
\caption{The difference between the conformal gravitational and photon horizon
in the high energy regime ($\ell H \gg 1$), between spatial hypersurfaces at
times A and B is illustrated. The future conformal photon horizon grows
linearly with conformal time, whereas the conformal gravitational horizon grow
as a power law. By flipping the diagram upside down, we can see that the past
gravitational horizon becomes larger for earlier starting time. However, as
argued in the text, the effect is not enough to solve the horizon problem
within the constraints of the five-dimensional theory.
\label{figure2}}
\end{figure}

Returning to our specific results, the difference between the gravitational and
photon horizons, may be enough to provide for some very interesting
physics. Specifically, gravitational effects on the brane propagate outside the
light cone, as illustrated in Figure~\ref{figure2}, due to the shortcuts
through the fifth dimension, forcing a redefinition of past and future causal
domains.  The communication of gravitational effects, both radiative and
non-radiative degrees of freedom, over length scales $r_g \gg r_\gamma$ beyond
the influence of fields on the brane has not yet been investigated. While we
cannot comment decisively, this seems to have an important bearing on at least
two problems: the initial conditions for inflation, and phase transitions in
the early Universe.

The inflaton must be homogeneous and potential-dominated over a region larger
than the horizon volume in order to initiate inflationary expansion. (See
\cite{VT00} for details.) If information about the gradients or inhomogeneities
in the inflaton field are carried by gravity, then correlations in the inflaton
can arise on length scales $r_g \gg r_\gamma$.  (Of course, it has already been
pointed out that correlations can exist on superhorizon scales \cite{W92}, but
the amplitude must decay \cite{RW96}.) The outcome depends on whether the
gravitational interaction leads to dissipation or amplification of
inhomogeneities.

The rate at which a phase transition proceeds in the expanding Universe, and
the formation of topological defects through the Kibble mechanism hinges on the
relative sizes of the correlation length of the order parameter and the causal
length scale. If information about the fields involved in the phase transition,
such as local fluctuations in the energy density, are carried by gravity, this
could affect the rate of the phase transition, and the rate at which
topological defects are formed. 

We also pause to mention that other analyses of phase transitions and
challenges to inflation (specifically, the flatness problem) have been carried
out in the context of the brane-world \cite{CKR00,DPDV00}.

Finally, we note that it is unlikely that shortcuts due to a local
gravitational distortion of the brane have an observable effect. Assuming that
the ratio $\ell / \lambda$ plays a similar role as $\ell H$ in determining the
size of the shortcut, where $\lambda = c (r^3 / G M)^{1/2} \sim 10^{13}$~cm for
the Earth, then the effect is  negligible.

\acknowledgements

We thank Dan Chung for useful comments. RRC thanks the Institut d'Astrophysique
de Paris for hospitality, where this work was initiated.


\begin{references}


\bibitem{HW}
P.~Horava and E.~Witten,
Nuc. Phys. {\bf B 475}, 94 (1996).

\bibitem{ADD98}
N.~Arkani-Hamed, S.~Dimopoulos, and G.~Dvali,
Phys. Lett. {\bf B429}, 263 (1998). 

\bibitem{RS99B} 
L.~Randall and R.~Sundrum, Phys. Rev. Lett. {\bf 83}, 4690 (1999).

\bibitem{ADD99}
N.~Arkani-Hamed, S.~Dimopoulos, and G.~Dvali, 
Phys. Rev. {\bf D 59}, 086004 (1999). 
 
\bibitem{BHKZ99}
V.~Barger, T.~Han, C.~Kao, R.~J.~Zhang,
Phys. Lett. {\bf B461}, 34 (1999). 

\bibitem{HS99}
L.~Hall, D.~Smith,
Phys. Rev. {\bf D60}, 085008 (1999). 


\bibitem{GRW98}
G.~Giudice, R.~Rattazzi, J.~Wells,
Nucl. Phys. {\bf B544}, 3 (1999). 

\bibitem{MPP98}
E.~Mirabelli, M.~Perelstein, M.~Peskin, 
Phys. Rev. Lett. {\bf 82}, 2236 (1999). 

\bibitem{H98}
J.~Hewett,
Phys. Rev. Lett. {\bf 82}, 4765 (1999). 

\bibitem{CHE00}
D.~Chung, H.~Davoudiasl, and L.~Everett,
{\it Experimental Probes of the Randall-Sundrum Infinite
Extra Dimension},  hep-ph/0010103.


\bibitem{gravtesta}
J.~C.~Long,  H.~W.Chan, and J.~C.~Price,
Nucl. Phys. {\bf B539}, 23 (1999).

\bibitem{gravtestb}
D.~E.~Krause and E.~Fischbach in {\it Testing General Relativity in Space: 
Gyroscopes, Clocks and Interferometers}, eds. by  Ammerzahl, Everitt, Hehl 
(Springer-Verlag, 2000); hep-ph/9912276. 

\bibitem{gravtestc}
C.~D.~Hoyle, {\it et al.}, 
Phys. Rev. Lett. {\bf 86}, 1418 (2001).


 \bibitem{BDL00}
P.~Bin\'{e}truy, C.~Deffayet, and D.~Langlois,
Nuc. Phys. {\bf B 565}, 269 (2000).

\bibitem{BDEL00}
P.~Bin\'{e}truy, C.~Deffayet, U.~Ellwanger, and D.~Langlois,
Phys. Lett. {\bf B 477}, 285 (2000).

\bibitem{cosmors}
C.~Cs\'aki, M.~Graesser, C.~Kolda, J.~Terning, 
Phys. Lett. {\bf B462}, 34 (1999);
J.~M.~Cline, C.~Grojean, G.~Servant,  
Phys. Rev. Lett. {\bf 83}, 4245 (1999).

\bibitem{CF99B}
D.~Chung and K.~Freese, 
Phys. Rev. {\bf D 62} 063513 (2000).

\bibitem{I00}
H.~Ishihara,  
Phys. Rev. Lett. {\bf 86}, 381 (2001). 

\bibitem{VT00}
T.~Vachaspati and M.~Trodden,
Phys. Rev. {\bf D 61}, 023502 (2000).
  
\bibitem{W92}
R.~M.~Wald, Gen. Rel. \& Grav. {\bf 24}, 1111 (1992).

\bibitem{RW96}
J.~Robinson and B.~D.~Wandelt,
Phys. Rev. {\bf D 53}, 618 (1996).

\bibitem{CKR00}
D.~Chung, E.~Kolb, and A.~Riotto, 
{\it Extra Dimensions Present a New Flatness Problem},
hep-ph/0008126.

\bibitem{DPDV00}
S.~C.~Davis, W.~B.~Perkins, A.-C.~Davis, I.~Vernon,
{\it Cosmological Phase Transitions in a Brane World},
hep-ph/0012223. 


\end{references}
\end{document}